\documentclass[conference]{IEEEtran}
\usepackage{cite}
\ifCLASSINFOpdf
  \usepackage[pdftex]{graphicx}
\else
\fi
\usepackage{amsmath}
\usepackage{mathtools}

\newcommand{\argmin}{\mathop{\rm argmin}\limits}

\usepackage{url}
\begin{document}
\title{
2D Sinc Interpolation-Based Fractional Delay and Doppler Estimation Using Time and Frequency Shifted Gaussian Pulses
}

\author{\IEEEauthorblockN{Yutaka Jitsumatsu}
\IEEEauthorblockA{Department of Information and Communication Engineering,\\
Tokyo Institute of Technology\\
Ookayama, Meguro-ku, Tokyo 145-8580\\
Email: jitsumatsu@ict.e.titech.ac.jp}
}


%Telephone: (800) 555--1212, Fax: (888) 555--1212}


\maketitle

\begin{abstract}
An accurate delay and Doppler estimation method for a radar system using time and frequency-shifted pulses with pseudo-random numbers is proposed. 
The ambiguity function of the transmitted signal has a strong peak at the origin and is close to zero if delay and Doppler are more than the inverses of the bandwidth and time-width. 
A two-dimensional (2D) sinc function gives a good approximation of the ambiguity function around the origin, 
by which fractional delay and Doppler are accurately estimated. 
\end{abstract}


% \fi
\IEEEpeerreviewmaketitle

\section{Introduction}
In recent years, the area of convergence between wireless communications and radar has attracted much attention ~\cite{jrac, dfrc, jcas}.
This is due to the high demand for radar in vehicle-to-vehicle communications to determine the relative position between vehicles for collision avoidance. In addition, due to the scarcity of the radio frequency band, devices are being developed that can use millimeter waves near the 60 GHz band for communication. In addition, since electromagnetic waves travel straight in this band, a simultaneous realization of wireless communications and radar is expected.

The Gabor Division Spread Spectrum (GDSS) system proposed by the author et al.~\cite{YutakaISSSTA2010, YutakaVTC2013fall, YutakaIRS2013, YutakaMACOM2013book} can be used as both communications and radar systems. A significant feature of the GDSS system is its ability to simultaneously use time-domain (TD) and frequency-domain (FD) spreading codes to detect the propagation delay and Doppler frequency between the radar system and an object, thereby estimating the distance and velocity of the moving target. 
Such a system is compatible with Orthogonal Time Frequency Space (OTFS)-based radar systems.

In an Orthogonal Time Frequency Space (OTFS) system~\cite{OTFS, OTFS-BITS}, data symbols
are located in the delay-Doppler (DD) domain. A single pilot symbol is located in $(0,0)$ coordinate in the delay-Doppler plane in OTFS communication system~\cite{Delay-Doppler-Communications, Pfadler2022}.
On the other hand, all $N\times M$ symbols are utilized for target detection in OTFS-based radar systems~\cite{Raviteja2019, Zhang2023Radar}. 
In this paper, we study a radar system using a time-frequency (TF) domain 
$N_{\rm t}<N$ times $N_{\rm f}<M$ symbols, where $N_{\rm f} \approx M/2$ and $N_{\rm t} \approx N/8$. 
We call radar system with such a coded pulse a GDSS radar~\cite{YutakaISSSTA2010, YutakaVTC2013fall, YutakaIRS2013, YutakaMACOM2013book}.

In our previous studies, the sampling frequency was set to several times faster than the Nyquist frequency of the bandwidth of the transmitted signal to make the delay estimation accurate. 
Increasing the sampling rate $n$ times makes the precision of the delay $1/n$ but increases the computation time. 
This paper proposes a method that samples at the Nyquist rate, estimates the delay coarsely, and then refines the delay by interpolating between samples.
The computational complexity is significantly reduced by searching only around the coarsely estimated signal.
A similar method of refinement of delay and Doppler has been proposed in~\cite{Zhang2023Radar} for OTFS-based radar systems. 
Unfortunately, the method in~\cite{Zhang2023Radar} does not work properly in GDSS radar systems  
mainly due to differences between the construction of signals in our method and~\cite{Zhang2023Radar}.  
The main contribution of this paper is that we show that the ambiguity function 
of the GDSS signal is accurately approximated by a two-dimensional (2D) sinc function.
Such an approximation enables us to realize a high-precision fractional delay and Doppler estimation method. 

\section{Transmitted signal and channel model}
This section gives the definition of the transmitted and received signals in the proposed radar system, which is based on our previous research on GDSS~\cite{YutakaISSSTA2010, YutakaVTC2013fall, YutakaIRS2013, YutakaMACOM2013book}.
In the GDSS system, the data symbol is multiplied by two-dimensional spread spectrum (SS) codes,
in the time-frequency domain. Such a modulation is similar to OTFS system. 
Unlike OTFS system, the GDSS system does not use the delay-Doppler domain signal processing to modulate and demodulate the data symbols. 

\begin{figure}
    \centering
    \includegraphics[width=\columnwidth]{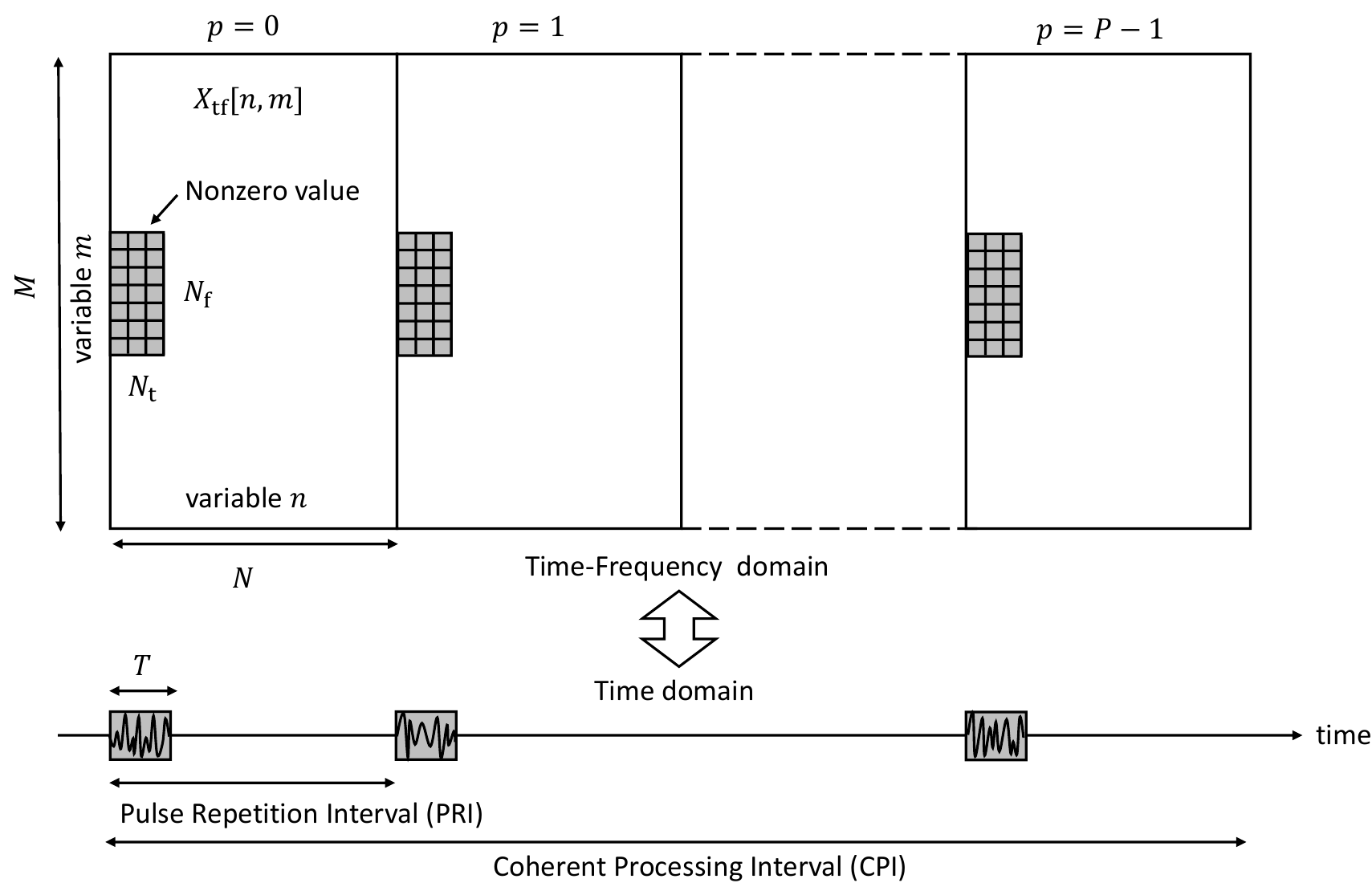}
    \caption{A construction of a GDSS-based pulsed radar signal}
    \label{fig:GDSS_radar}
\end{figure}

The construction of the transmitted signal of GDSS-based pulse radar is shown in Fig.~\ref{fig:GDSS_radar}. 
A coded pulse $s(t)$ is transmitted repeatedly. Here, we concentrate on one pulse repetition interval (PRI).
We assume a single-antenna pulsed radar system, where the transmitter and the receiver share the antenna.
A pulse with a pulse width $T=N_{\rm t} T_c$ is transmitted and reflected by the object, where $N_{\rm t} < N$. 
The radar system receives signals except when it is transmitting pulses.
PRI is typically ten times more than the pulse duration $T$.

The continuous-time received signal in the absence of noise is
\begin{align}
    r(t) = \alpha s(t-t_{\rm d}) {\rm e}^{{\rm i} 2\pi f_{\rm D} t},
    \label{r(t)}
\end{align}
where $\alpha$ is a complex attenuation factor, $t_{\rm d} =(2R)/c$ is a round time, 
where $R$ is the range of the target and $c$ is the speed of light,
$f_{\rm D} = (v/c) f_{c}$ is the Doppler frequency, where $v$ is the velocity of the target
$f_{\rm c}$ is the carrier frequency. 
Suppose that the propagation delay satisfies $ N_t T_c \le t_{\rm d} \le (N-N_t) T_c $ 
so that we can avoid receiving the reflected pulse during the transmission 
and detect the target within one PRI.
Assume that we only use $N_{\rm f}<M$ subcarriers for constructing the transmitted signal. 
Data symbols can be overlaid with high-frequency components within a $[0,T]$ time width. 
The focus of this paper is on accurate estimation of delay and Doppler, and the simultaneous realization of radar and communications is a future challenge.

%The difference between continuous-time and discrete-time models is important, especially in the fine estimation of delay and Doppler. 
The transmitted signal is defined in continuous time as 
%\footnote{A frequency shift with phase $-\pi n m T_c F_c $ for each element $(n,m)$ was introduced in our previous studies, which is omitted in this paper for simplicity.} 
\begin{align}
    s(t) = \sum_{n=0}^{ N_{\rm t} -1} \sum_{m = - N_{\rm f}/2}^{ N_{\rm f}/2 -1} X_{\rm tf}[n,m] g_{\rm tx}\left(\textstyle \frac{t}{T_c} - n \right)
    e^{ {\rm i}  2\pi m F_c t},
    \label{s(t)}
\end{align}
where $\rm i$ denotes an imaginary unit,
$X_{\rm tf}[n,m]$ is transmitted symbols, or two-dimensional spreading codes in GDSS systems. 
$g_{\rm tx}(t)$ is a transmitter shaping waveform, 
$T_c$ is a pulse interval and $F_c = 1/T_c$ is subcarrier spacing. 
We usually let $m$ run from $0$ to $N_{\rm f}-1$. Here, in order to make $s(t)$ to use a low-frequency band, we made $m$ to run from $-N_{\rm f}/2$ to $N_{\rm f}/2-1$, when $M$ is an even number. 
A Gaussian waveform $g_{\rm tx}(t) = \sqrt[4]{2} \exp( -\pi t^2)$ is used in GDSS systems. 
All $N_{\rm t} \times N_{\rm f}$ symbols are used for delay-Doppler estimation.

Suppose that $s(t)$ as well as the received signal $r(t)$ are perfectly bandlimited.
Then, the sampling theorem guarantees that the continuous-time signal is completely recovered from its samples if the sampling rate is faster than the Nyquist rate. 
Assume the maximum frequencies of $s(t)$ and $r(t)$ are lower than $MF_c/2$ Hz so that the Nyquist sampling interval is $T_s = 1/(MF_c)$. 
Let the discrete-time version of (\ref{s(t)}) be $s[j ] = s(j T_s)$, 
expressed by 
\begin{align}
    s[j] = \frac{1}{M}
    \sum_{n=0}^{N_{\rm t}-1} \sum_{m=-N_{\rm f}/2}^{N_{\rm f}/2-1} 
    X_{\rm tf}[n,m] g_{\rm tx}[ j - n M ] W_{M}^{-mj},
    \label{s[j]}
\end{align}
where $W_{M} ={\rm e}^{- \frac{2\pi {\rm i} }{M} }$ is a twiddle factor of
$M$-point DFT and $g_{\rm tx}[j] = g_{\rm tx}(jT_s)$ are samples of a shaping pulse\footnote{Practically, $g[j]$ with negligibly small value can be truncated and $3M$ is sufficient for the number of samples for $g[j]$.}.
The discrete-time signal $s[j]$ is time-limited. 
%Let the length of the signal be $L_s$, which is $(N_{\rm t}-1) M$ plus the length of the filter $g_{\rm tx}[j]$. 
The transmitted signal is a Digital-to-Analog (DA) converted signal of (\ref{s[j]}) which is identical to (\ref{s(t)}), provided that the assumptions for the sampling theorem are satisfied. Note that a rectangular pulse for $g_{\rm tx}(t)$ does not satisfy the condition. 
By definition, the index $j$ in (\ref{s[j]}) runs from $0$ to $NM-1$. However, 
the indices with nonzero $s[j]$ are limited to the first $N_{\rm t} M$ samples with the tail of the Gaussian pulse.
Let $L = (N_{\rm t} + 2) M $ be the number of samples with nonzero $s[j]$.

The transmitted and received signals are discrete time, 
but the delay and Doppler $t_{\rm d}, f_{\rm D}$ are real values.
The input-output relation of the channel with a fractional delay and Doppler in~\cite{Zhang2023Radar} is expressed in the delay-Doppler domain, which is derived by~\cite{Wei2021}.
This paper uses the simpler model below.
Assume $r[i] = \int r(t) h_{\rm ad}(\frac{t}{T_s}-i) dt$ and
$s(t) = \sum_j s[j] h_{\rm da} ( \frac{t}{T_s} - j)$, where $h_{\rm da}(t)$ and $h_{\rm ad}(t)$ are, respectively, a DA and an Analog-to-Digital (AD) filters used in the transmitter and the receiver.
These filters should not be confused with $g_{\rm tx}(t)$ and $g_{\rm rx}(t)$, which are the prototypes of the digital filters $g_{\rm tx}[j]$ and $g_{\rm rx}[j]$. While $h_{\rm tx}(t)$ and $h_{\rm rx}(t)$ are analog filters. 
Then, 
the discrete-time received signal in the presence of noise is given by
\begin{align}
    r[i] = \sum_{j=0}^{L-1} H_{ij}(t_{\rm d}, f_{\rm D}) s[j] + \eta[i],
\end{align}
where $\eta[i]$ is a complex white Gaussian noise and 
\begin{align}
    & H_{ij}(t_{\rm d},f_{\rm D}) \notag\\
    & = \int_{-\infty}^{\infty} \textstyle 
    h_{\rm da} \left( \frac{t - t_{\rm d} }{T_s} - j \right)
    h_{\rm ad}^* \left( \frac{t}{T_s} - i \right) {\rm e}^{{\rm i}2\pi f_{\rm D} t}dt.
\end{align}

For simplicity, assume that $h_{\rm da}(t)$ and $h_{\rm ad}(t)$ are ideal sinc functions.
Then  we have
\begin{align} 
    H_{ij}(t_{\rm d}, f_{\rm D}) &= \textstyle 
    T_s [ \frac{1}{T_{\rm s}} -|f_{\rm D}|]^+ {\rm e}^{ {\rm i} \pi (t_{\rm d} + (m+n) T_s ) f_{\rm D} } \notag\\
    & \textstyle \quad {\rm sinc} \left \{ (\frac{1}{T_s} - |f_{\rm D}|) ( t_{\rm d} + 
    (m-n)T_{\rm s} ) \right \} ,
\end{align}
where $[x]^+ = \max \{ 0, x\}$. 

For a pulse radar application, the transmitted pulse must be localized in time domain. 
Therefore, the signal is better to be designed in the time-frequency domain. 
In the OTFS system, on the other hand, all $N\times M$ positions of $X_{\rm tf}[n,m]$ are used
and data symbols are defined in delay-Doppler domain $X[k,\ell]$. The relation between 
$X[k,\ell]$ and $ X_{\rm tf}[n,m]$ is given by
\begin{align}
    X[k,\ell] = \frac{1}{M} 
    \sum_{n=0}^{N-1} \sum_{m=-M/2}^{M/2-1} 
    X_{\rm tf}[n,m]  
    W_{N}^{-nk} W_{M}^{ m \ell}. 
\end{align}
$X[k,\ell]$ is periodic in both $k$ and $\ell$
with periods $N$ and $M$ respectively.

\section{2D sinc approximation}
The proposed method consists of two steps: 
the first step is a coarse estimation and the second is a fine estimation.
Let $\ell_{\rm d}$ and $k_{\rm D}$ are nearest integer in $T_s$ and $\Delta f = \frac{1}{NMT_s}$
and $\epsilon_{\rm t}$ and $\epsilon_{\rm f}$ are fraction parts of the delay and the Doppler
with $|\epsilon_{\rm t}|, |\epsilon_{\rm f}|\le 1/2$, 
i.e.,
$$
t_{\rm d} = (\ell_{\rm d} + \epsilon_{\rm t} ) T_s, \quad
f_{\rm D} = ( k_{\rm D} + \epsilon_{\rm f} ) \Delta f. 
$$

The ambiguity function plays a central role in radar systems~\cite{Woodward}.
The ambiguity function between an $x(t)$ and a $y(t)$ is defined by
\begin{align}
    A_{x,y}(\tau, \nu) =\int x(t) y^*(t-\tau) {\rm e}^{-2\pi {\rm i} \nu t} dt. 
    \label{Ars(tau,nu)}
\end{align}
%In the receiver, samples of $r(t)$ and $s(t)$ are stored.
In the proposed system, the discrete-time transmitted signal $s[j]$ is stored, and the received signal sampled at each $T_s$ is denoted by $r[j]$. 
Then, we first calculate the discrete-time ambiguity function.
The discrete-time ambiguity function is defined by
%    Calculate an FFT-based ambiguity function between the discrete-time $r[j]$ and $s[j]$, 
\begin{align}
    A_{r,s}[\ell, k] = \sum_{j=0}^{NM-1} r[j] s^*[j-\ell] W_{NM}^{k j}. 
    \label{Ars[l,k]}
\end{align}
This calculation is done efficiently using the Fast Fourier Transform (FFT).
The discrete-time ambiguity function is equal to the continuous-time one with $\tau = \ell T_s$ and $\nu = k \Delta f$.

Let $(\hat \ell_{\rm d}, \hat k_{\rm D})$ be the index of the maximum absolute value of $A_{r,s}[\ell,k]$. 
Then, we find the $( ( \hat \ell_{\rm d} + \varepsilon_{\rm t}) T_s , ( \hat k_{\rm D} + \epsilon_{\rm f}) \Delta f )$ 
that takes the maximum absolute value of $A_{r,s}(\tau,\nu)$, by interpolating the discrete-time ambiguity function. 

Substituting (\ref{r(t)}) into (\ref{Ars(tau,nu)}) with $x=r$ and $y=s$, we obtain 
\begin{align}
    A_{r,s}(\tau, \nu) = \alpha {\rm e}^{-2\pi {\rm i} (\nu - f_{\rm D}) t_{\rm d} }
    A_{s,s}(\tau - t_{\rm d}, \nu - f_{\rm D}) 
    \label{Ars_and_Ass}
\end{align}

From (\ref{Ars[l,k]}) and (\ref{Ars_and_Ass}), we have
\begin{align}
    A_{r,s}[\ell, k ] &=\alpha \cdot {\rm e}^{-2\pi {\rm i} (\nu - t_{\rm d}) t_{\rm d} }\notag\\
    & \quad
    \cdot 
    A_{s,s}( (\ell-\ell_{\rm d} - \epsilon_{\rm t}) T_s , (k-k_{\rm D} - \epsilon_{\rm f} ) \Delta f )
    \label{Ars[l,k]-2}
\end{align}
Thus, first, find $\ell$, and $k$ that attain the maximum absolute value of $A_{rs}[\ell, k]$.
Suppose that they are the integer parts of the delay and Doppler.
Then we can estimate the remaining $\epsilon_{\rm t}$ and $\epsilon_{\rm f}$ by optimal matching.

If the values of $A_{s,s}(\tau,\nu)$ are available for any $ | \tau |< 1/(N_{\rm f} F_c)$ 
and $| \nu | < 1/(N_{\rm t} T_c)$ with high precision, we can find $\epsilon_{\rm t}, \epsilon_{\rm f}$ 
by two-dimensional (2D) matching. 
However, the value of $A_{ss}(\tau, \nu)$ depends on the choice of $X_{\rm tf}[n,m]$ and 
computing $A_{s,s}( \tau, \nu)$ with such high precision is time-consuming.
To solve this problem, we propose to approximate $A_{s,s}( \tau, \nu)$.
The fundamental properties of the auto-ambiguity function $A_{s,s}(\tau,\nu)$ are the followings:
\begin{itemize}
    \item The bandwidth of $s(t)$ is approximately $N_{\rm f}F_c$. Hence, the time resolution is
    $1/(N_{\rm f}F_c)$. The time resolution divided by the sampling interval is $M/N_{\rm f}$. 
    \item The length of $s(t)$ with nonzero value is $ N_{\rm t}T_c$.
    Hence, the frequency resolution is $1/ ( N_{\rm t}T_c ) $.
    The frequency resolution divided by the frequency bin is $N/N_{\rm t}$.
    \item The above two properties imply that 
    $A_{s,s}(\tau, \nu)$ takes a large value if $ | \tau |< 1/(N_{\rm f} F_c)$
    and $| \nu | < 1/(N_{\rm t} T_c)$. 
\end{itemize}

\begin{figure}
    \centering
    \includegraphics[width=\columnwidth]{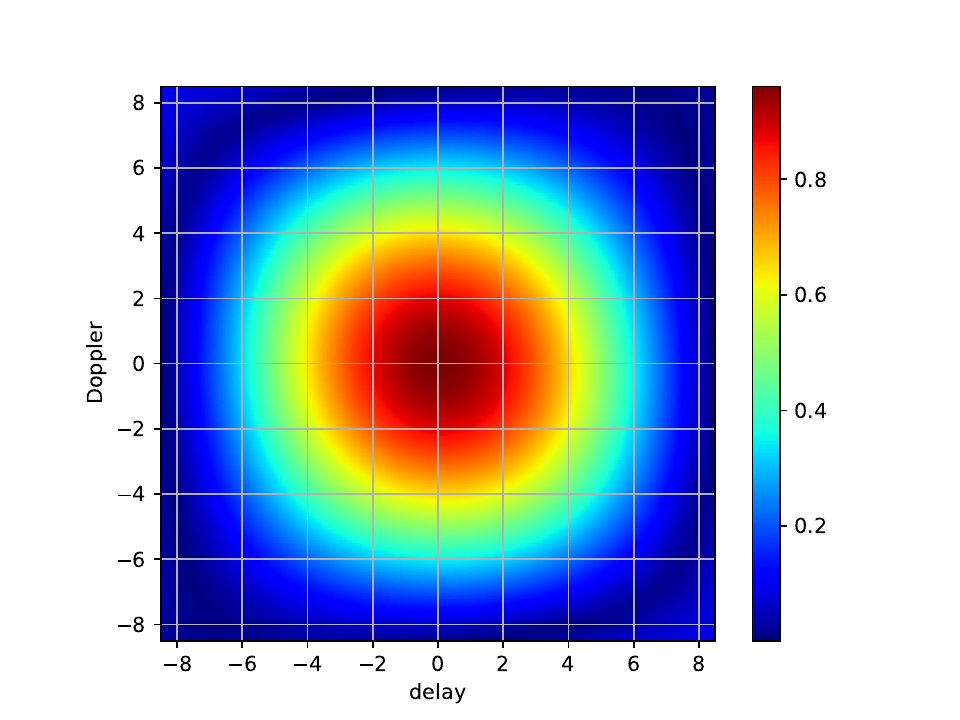}
    \caption{An example of the absolute value of the ambiguity function of $s(t)$
    around the origin} 
    \label{fig:Contour_plot}
\end{figure}

\begin{figure}
    \centering
    \includegraphics[width=0.47\columnwidth]{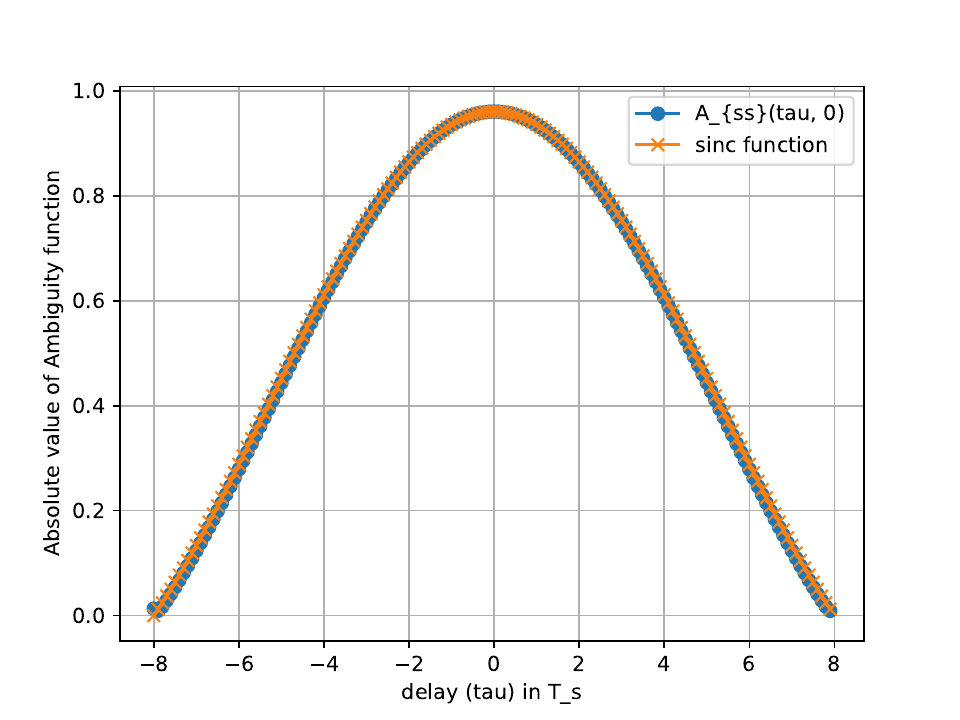} \ 
    \includegraphics[width=0.47\columnwidth]{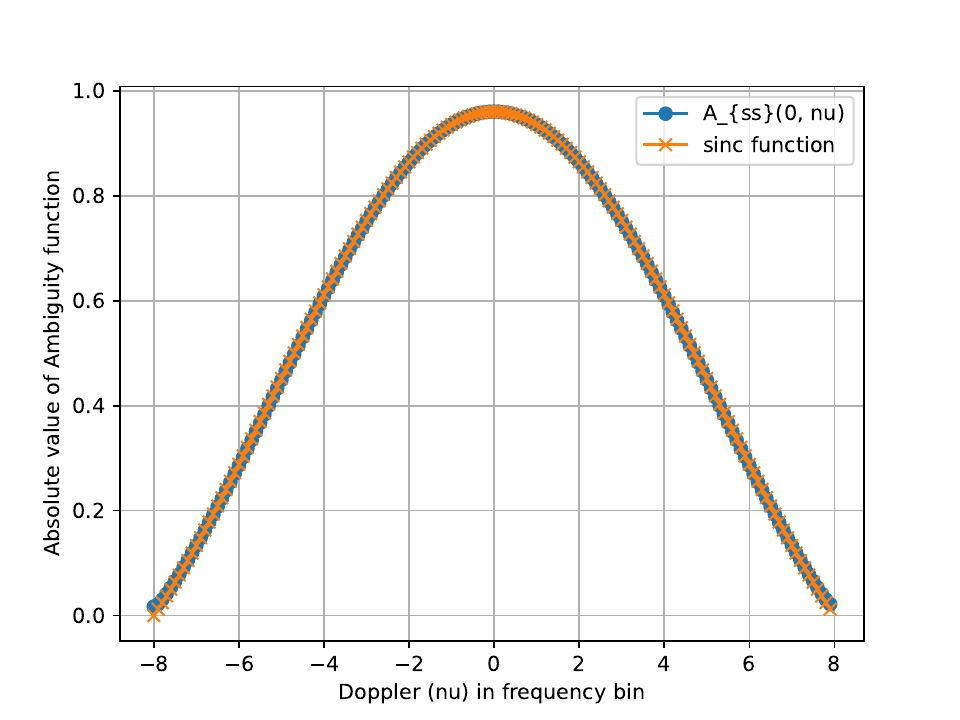}
    \caption{The absolute values of the ambiguity function along $\tau$ and $\nu$ axes of Fig.\ref{fig:Contour_plot}.}
    \label{fig:ambiguityfunction_time_and_frequency}
\end{figure}

We observe that if $N_t$ and $N_f$ are sufficiently large, then 
the shape of $A_{ss}(\tau, \nu)$ in $|\tau|< 1/(N_{\rm f} F_c) $ and 
$| \nu  |< 1/(N_{\rm t} T_c) $ is almost irrespective of $X_{\rm tf}[n,m]$.
An example of $|A_{r,s}(\tau, \nu)|$ for $N_t=F_f=8$, $N=M=64$ 
with 
\begin{align}
\scriptsize
X_{\rm tf}[n,m]=\begin{bmatrix*}[r]
-1&  1& -1&  1& -1&  1& -1& -1\\
-1& -1& -1&  1&  1& -1&  1&  1\\
-1&  1&  1& -1& -1& -1&  1&  1\\
-1& -1&  1&  1&  1& -1& -1&  1\\
-1&  1&  1&  1& -1& -1& -1& -1\\
-1&  1&  1& -1& -1& -1&  1&  1\\
-1& -1&  1&  1&  1& -1& -1&  1\\
 1&  1&  1&  1& -1&  1&  1& -1 
\end{bmatrix*}
\label{good_code}
\end{align}
is shown in Fig.~\ref{fig:Contour_plot}, where the horizontal axis is  
$\tau$ in $T_s$ and the vertical axis is $\nu$ in $\Delta f$. 
The ambiguity function is normalized to be $A_{s,s}(0,0)=1$.
%This figure is centered at $(\ell,k) = (0,0)$. 
This figure shows that the contours of the ambiguity function are concentric. 
Rigorously speaking, the contours are not perfect circles.
In Fig.~\ref{fig:ambiguityfunction_time_and_frequency}, one-dimensional plots of $A_{ss}(\tau,\nu)$ 
cut by the $\tau$ and $\nu$ axes are shown. 
The curves in this figure coincide with the sinc function $\frac{\sin \pi x }{ \pi x }$  
of $x \in [-1,1]$ with extremely high accuracy.
This fact enables us to estimate the fractional delay and Doppler accurately.


As a reason why the central part of the ambiguity function can be approximated by the sinc function, 
consider the expectation value taken for a random $X_{\rm tf}[n,m]$ for the auto-ambiguity function. 
Define 
\begin{align}
    \bar A(\tau,\nu) = {\rm E} [ A_{s,s} (\tau,\nu)]
\end{align}
where the expectation is taken over the random $X_{\rm tf}[n,m]$.
We have 
\begin{align}
    \bar A(\tau, \nu) 
    & = %\frac{1}{M^2}
    \sum_{n = 0}^{N_t-1}
    \sum_{m = - N_f/2}^{N_f/2-1}
    {\rm e}^{j2\pi (m F_c \tau - n T_c \nu )}
    A_{g_{\rm tx}, g_{\rm tx}} (\tau, \nu).
\end{align}    
The summations with respect to $n$ and $m$ yield Dirichlet kernels, i.e.,
$\sin ( \pi N_t T_c \nu ) / \sin (\pi T_c \nu) $ 
and $\sin ( \pi N_f F_c \tau ) / \sin (\pi F_c \tau) $ with a phase.
For $|\nu| < 1/(N_{\rm t} T_c) $ and $|\tau| < 1/ (N_{\rm f} F_c) $, 
we have 
\begin{align}
\sin ( \pi N_{\rm t} T_c \nu ) / \sin (\pi T_c \nu) &\approx N_{\rm t} {\rm sinc} ( N_{\rm t} T_c \nu ), \\
\sin ( \pi N_{\rm f} F_c \tau ) / \sin (\pi F_c \tau) &\approx N_{\rm f} {\rm sinc} ( N_{\rm f} F_c \tau ) , 
\end{align}
and $A_{g_{\rm tx}, g_{\rm tx}}(\tau,\nu) \approx 1$ if $|\tau|< 1/(N_{\rm f} F_c)$ and $|\nu| < 1/(N_{\rm t} T_c)$.
Then, we have the following approximation. 
\begin{align}
    \bar A(\tau, \nu) 
    &\approx 
    N_{\rm t} N_{\rm f} {\rm e}^{j\pi ( F_c \tau - (N_{\rm t}-1) T_c \nu ) } \notag \\
    & \quad \cdot {\rm sinc}( N_{\rm f} F_c \tau ) \cdot {\rm sinc}( N_{\rm t} T_c \nu ) 
   \notag  
    \\
    & \text{for} \, |\tau|<1/(N_{\rm f} F_c) \text{ and }\, |\nu| < 1/(N_{\rm t} T_c) \label{double-sinc}
\end{align}

\begin{figure}
    \centering
    \includegraphics[width=\columnwidth]{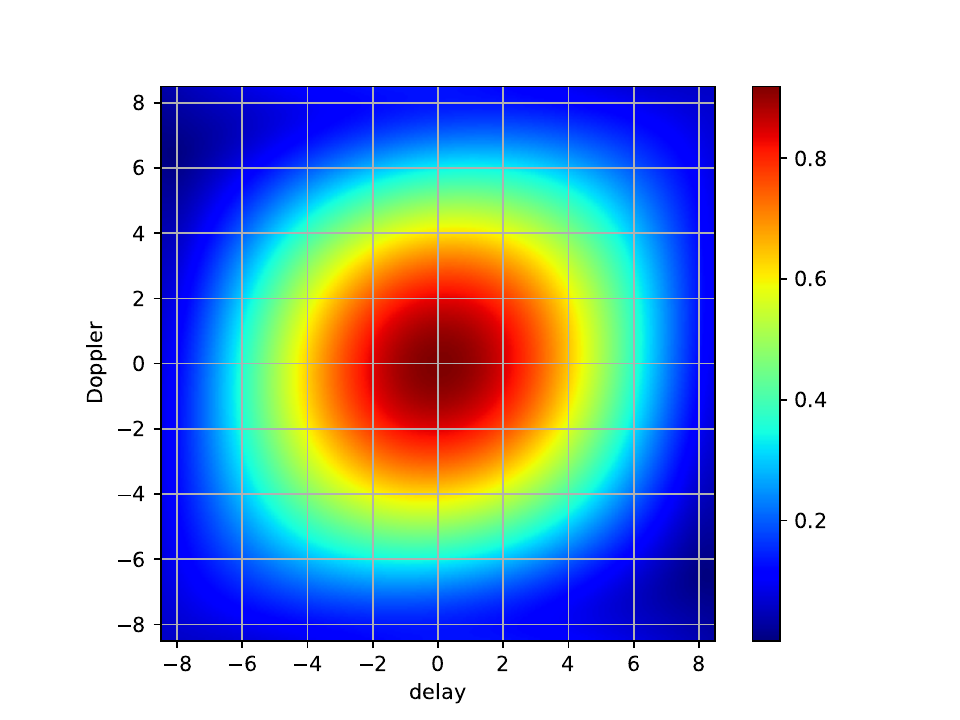}
    \caption{An undesirable example of the absolute value of the ambiguity function of $s(t)$
    around the origin.}
    \label{fig:Contour_plot_bad}
\end{figure}

\begin{figure}
    \centering
    \includegraphics[width=0.47\columnwidth]{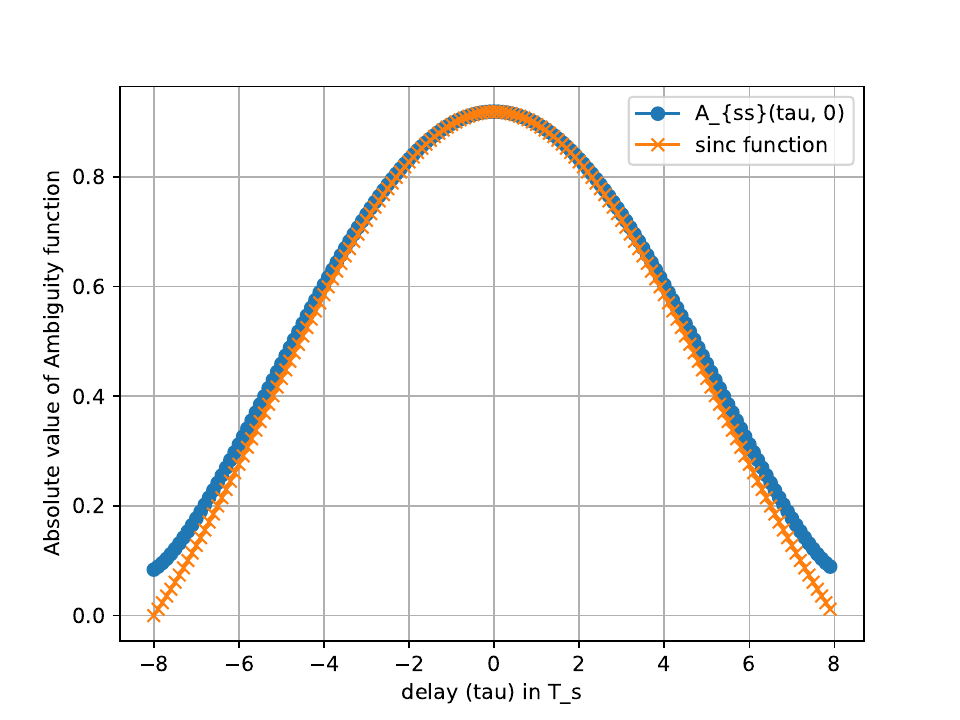} \ 
    \includegraphics[width=0.47\columnwidth]{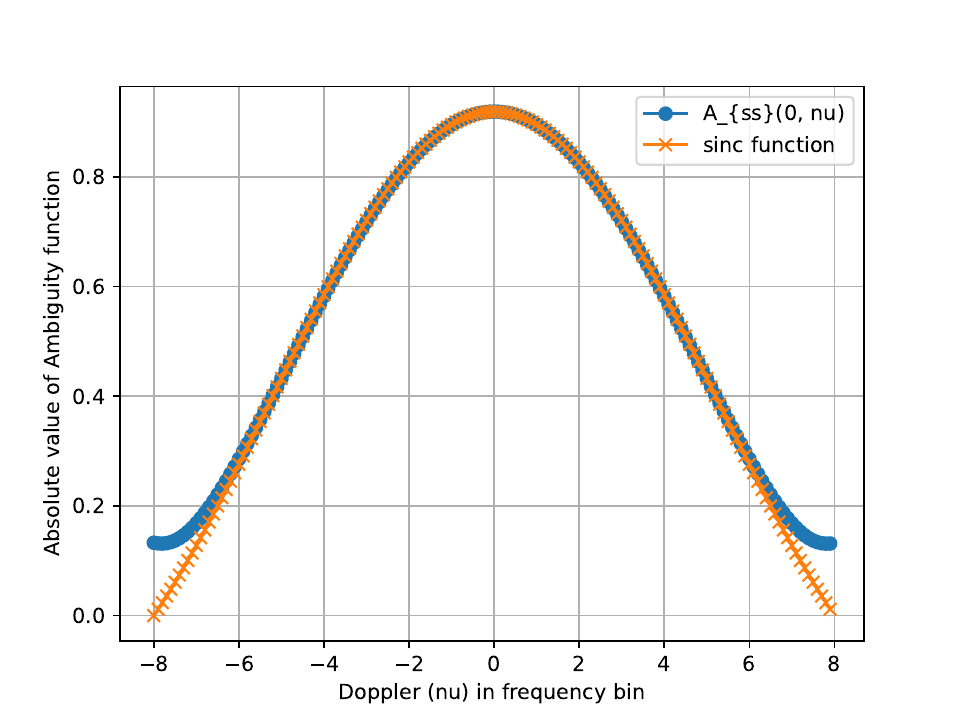}
    \caption{One-dimensional plots of the absolute value of the ambiguity function cut by the 
    $\nu$ and $\tau$ axes of Fig.~\ref{fig:Contour_plot_bad}.     
    }
    \label{fig:ambiguityfunction_time_and_frequency_bad}
\end{figure}

Note that Fig.~\ref{fig:Contour_plot} is not the ambiguity function of a pulse shape $g(t)$ but of $s(t)$ in (\ref{s(t)}) and therefore it depends on the choice of $X_{\rm tf}[n,m]$. 
We should not use a code $X_{\rm tf}[n,m]$ if its ambiguity function has a large difference from the 2D sinc function.
The absolute value of the ambiguity function with a code  
\begin{align}
    \scriptsize
    X_{\rm tf}[n,m] =\begin{bmatrix*}[r]
  -1 & -1 &  1 & -1 & -1 & -1 &  1 &  1\\
   1 &  1 &  1 &  1 &  1 & -1 &  1 &  1\\
  -1 & -1 & -1 & -1 & -1 &  1 & -1 & -1\\
   1 & -1 & -1 & -1 &  1 &  1 &  1 & -1\\
  -1 & -1 &  1 &  1 &  1 &  1 & -1 & -1\\
   1 &  1 & -1 &  1 &  1 & -1 &  1 &  1\\
   1 & -1 &  1 & -1 &  1 & -1 &  1 &  1\\
  -1 &  1 &  1 & -1 & -1 & -1 &  1 &  1\\
  \end{bmatrix*}
\end{align}
is shown in Fig.~\ref{fig:Contour_plot_bad}. 
It can be seen that the contour curves are skewed.
Fig.\ref{fig:ambiguityfunction_time_and_frequency_bad} shows one-dimensional plots
and that the curves along the $\tau$ and $\nu$ axes deviate significantly from the sinc function.
Hence we reject to use such a code.

\section{Proposed Method}

Now we state the proposed method for fractional delay and Doppler estimation based on the 2D sinc interpolation of the ambiguity function, as follows:

{\bf Algorithm 1 (2D Sinc interpolation)}
\begin{enumerate}
    \item Generate $X_{\rm tf}[n,m]$ randomly. Compute the transmitted signal (\ref{s[j]}).
     If the absolute ambiguity function $A_{ss}[\ell, k]$ is well approximated by 2D sinc function, fix the random numbers.
     Otherwise generate $X_{\rm tf}[n,m]$ again. 
    \item The system sends the first pulse $s[j]$ for $j =0,1,\ldots, L-1$. 
    The samples of the received signal are $r[j]$ for $j=L_, L_+1, NM-1$. 
    Let $r[j]=0$ for $j<L$.

%    Then find the maximum value of its absolute value, where $s[j-\ell]$ is basically a non-cyclic shift of $s[j]$ but the effect of replacing it with a cyclic shift is negligible.
    Let $\theta>0$ be a prescribed threshold value and list up 
    the $(\hat \ell_{\rm d}, \hat k_{\rm D}) = (\ell,k)$ satisfying $ | A_{r,s}[ \ell, k] | > \theta $. 
    If there is no such $(\ell, k)$, we proceed to the next frame.
%    Or simply let $(\hat \ell_{\rm d}, \hat k_{\rm D}) = \arg \max_{\ell, k} | A[\ell, k] |$.
    \item 
    For each $(\hat \ell_{\rm d}, \hat k_{\rm D})$ obtained by the first step, 
    a fine estimation based on a 2D sinc function approximation is performed as follows:
    \begin{align}
        &(\hat \alpha, \hat \epsilon_{\rm t}, \hat \epsilon_{\rm f})\notag\\
        &=
        \argmin_{\alpha\ge 0, \epsilon_{\rm t}, 
        \epsilon_{\rm f}}
        \sum_{ | \ell | \le  \lfloor \frac{M}{N_{\rm f}} \rfloor}
        \sum_{ | k | \le \lfloor \frac{N}{N_{\rm t}} \rfloor} \Big\{  \lvert A_{r,s}[\hat \ell_{\rm d} + \ell, \hat k_{\rm D} + k ] \rvert \notag \\
        & \quad 
        - 
          \alpha \lvert \bar{A} ( (\ell - \epsilon_{\rm t}) T_s , (k - \epsilon_{\rm f})\Delta f ) \rvert \Big\}^2 
          \label{argmin_epsilon_t_f}
    \end{align}
    where the range of $\epsilon_{\rm t}$ and $\epsilon_{\rm f}$ is $[-\frac12,\frac12]$.
    Then the fine-tuned estimate of delay and Doppler are
    \begin{align}
        \hat t_{\rm d} = (\hat \ell_{\rm d}+\hat \epsilon_{\rm t}) T_s,\quad
        \hat f_{\rm D} = ( \hat k_{\rm D} + \hat \epsilon_{\rm f}) \Delta f.
        \label{fine_estimation}
    \end{align}
\end{enumerate}

%The computation of (\ref{argmin_epsilon_t_f}) is performed using an optimization package in Python.
To obtain the solution to the minimization problem of (\ref{argmin_epsilon_t_f}),
a nonlinear optimization must be performed.
In some cases, simple estimation may be preferable to implementing complex algorithms.
To deal with such cases, we approximate the ambiguity function around the origin as a quadratic function. 
We approximate the absolute value of the ambiguity function as
\begin{align}
    \lvert \bar{A}(\tau,\nu) \rvert = - a \left( \frac{\tau}{T_s} \right) ^2 -b \left( \frac{\nu}{\Delta f} \right)^2 + c, 
\end{align}
where $a$, $b$ and $c$ are unknown parameters.
From (\ref{Ars[l,k]-2}) and the above approximation,
we have
\begin{align}
   & | A_{r,s}[\ell_{\rm d} + \ell, k_{\rm D}+k] | \notag\\
    &\approx - a (\ell-\epsilon_{\rm t})^2 - b  (k-\epsilon_{\rm f})^2 + c
    \label{abs_Ars}
\end{align}

We assume $\hat \ell_{\rm d} = \ell_{\rm d}$ and 
$\hat k_{\rm D} = k_{\rm D}$ and use five points of 
(\ref{abs_Ars})
with $(\ell, k) = (0,0), (\pm 1, 0), (0, \pm1) $.
Assuming that (\ref{abs_Ars}) holds with exact equality and solving the simultaneous equations with respect to $a, b, c, \epsilon_{\rm t}, \epsilon_{\rm f}$, the following fractional delay and Doppler estimates are obtained:
\begin{align}
    \hat \epsilon_{\rm t} & = \frac{|A_{rs}[\ell_{\rm d} + 1, k_{\rm D}] | - |A_{rs}[\ell_{\rm d} -1 , k_{\rm D}] | }{ 4 |A_{rs}[\ell_{\rm d}, k_{\rm D}] | - 2  |A_{rs}[\ell_{\rm d}+1, k_{\rm D}] |  -2 |A_{rs}[\ell_{\rm d}-1, k_{\rm D}] | } 
    \label{epsilon_t}\\
    \hat \epsilon_{\rm f} &= \frac{|A_{rs}[\ell_{\rm d} , k_{\rm D} +1 ] | - |A_{rs}[\ell_{\rm d} , k_{\rm D} -1 ] | }{ 4 |A_{rs}[\ell_{\rm d}, k_{\rm D}] | - 2  |A_{rs}[\ell_{\rm d}, k_{\rm D}+1] |  -2 |A_{rs}[\ell_{\rm d}, k_{\rm D}-1] | } 
    \label{epsilon_f}
\end{align}

{\bf Algorithm 2 (Quadratic Interpolation)}
The first and the second steps are the same as Algorithm 1.
(\ref{argmin_epsilon_t_f}) is replaced by 
(\ref{epsilon_t}) and (\ref{epsilon_f}) to determine $\hat \epsilon_{\rm t}$ and $\hat \epsilon_{\rm f}$.
Then compute (\ref{fine_estimation}) to obtain a fine delay and Doppler estimation.

\section{Simulation Results}

\begin{figure}
    \centering
    \includegraphics[width=4cm]{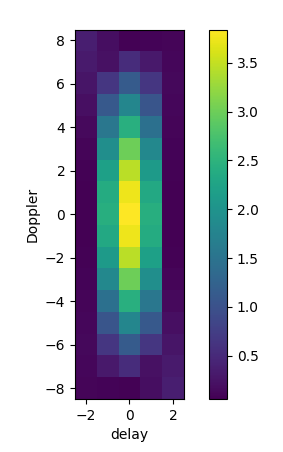}
    \caption{The discrete-time ambiguity function of the pulse}
    \label{fig:A_ss}
\end{figure}

\begin{figure}
    \centering
    \includegraphics[width=\columnwidth]{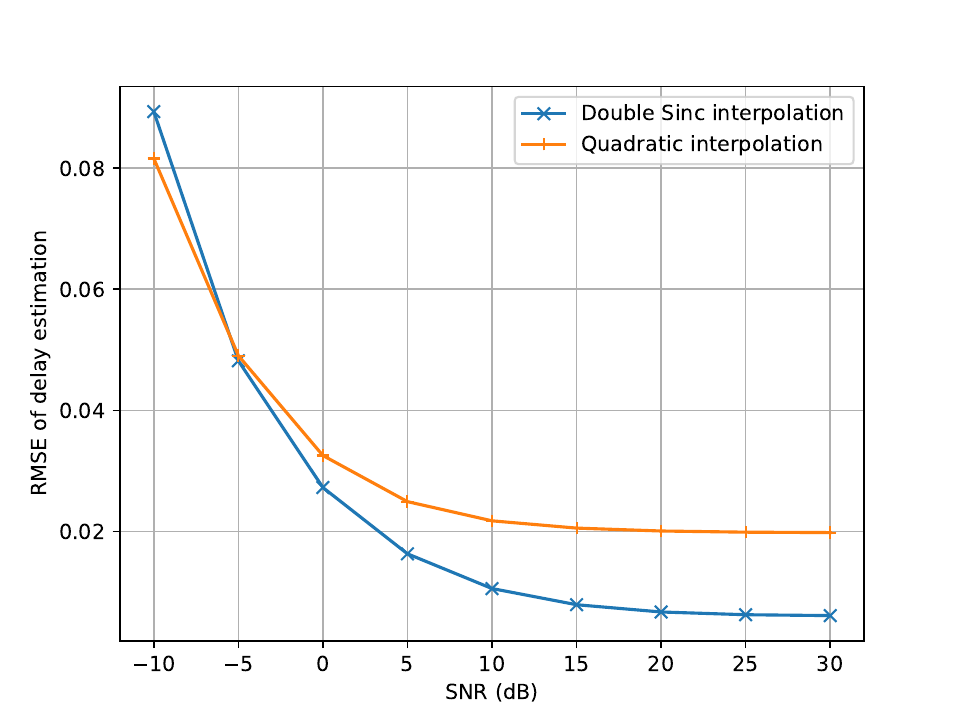}
    \caption{The mean square error of the estimated delay of the proposed method and the conventional method}
    \label{fig:RMSE_delay}
\end{figure}

\begin{figure}
    \centering
    \includegraphics[width=\columnwidth]{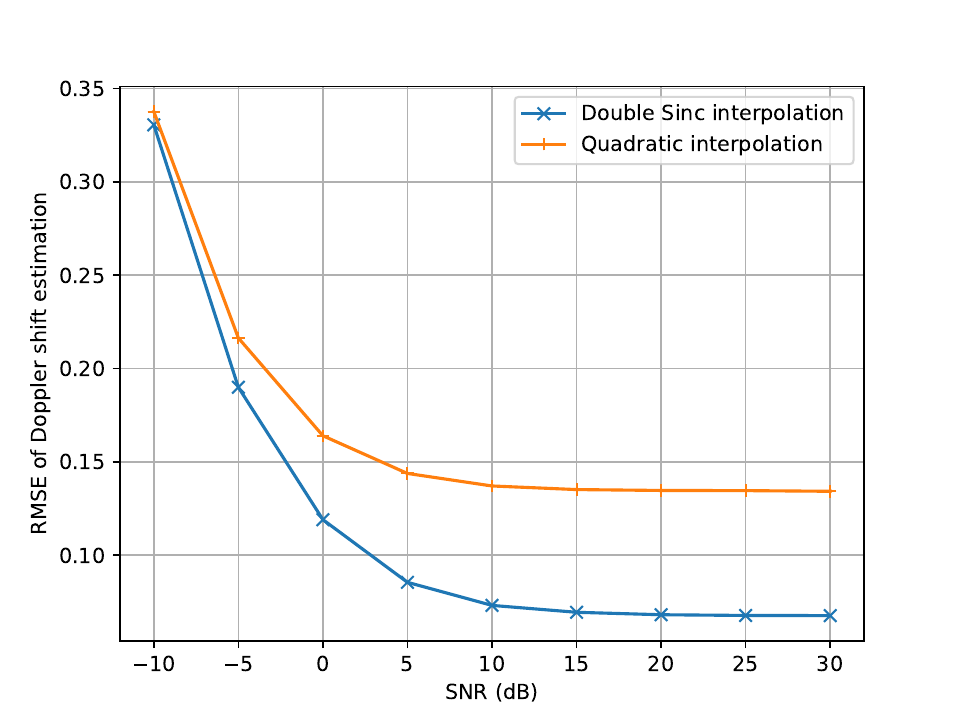}
    \caption{The mean square error of Doppler frequency }
    \label{fig:RMSE_Doppler}
\end{figure}

Algorithm 1 is implemented by a python code\footnote{The code is available at \url{https://github.com/jitumatu/}}.
The argmin in (\ref{argmin_epsilon_t_f}) is executed by scipy.optimize.minimize with Broyden–Fletcher–Goldfarb–Shanno (BFGS) algorithm. The initial guess is set as $(\alpha, \epsilon_{\rm t}, \epsilon_{\rm f} ) = (1,0,0)$.
The simulation were performed with $N_t = N_f =8,$ $N=64, M=16$.
A single pulse for one frame is transmitted. 
The discrete delay and Doppler ambiguity function of the signal at $|\ell| \le \frac{M}{N_f} =2$
and $|k| \le \frac{N}{N_t} = 8$ is shown in Fig.~\ref{fig:A_ss}. 
Fractional delay and Doppler are estimated by the proposed method.
The simulation results are shown in Figs.~\ref{fig:RMSE_delay} and~\ref{fig:RMSE_Doppler}.
The horizontal axis shows the SNR and the vertical axis shows the
root mean square error (RMSE) of the delay and Doppler.
The simulation is performed $K=10^3$ times. 
SNR is calculated by $\sum_{i=0}^{NM-1} | s[i] |^2 / (NM\sigma^2)$,
where $\sigma^2 $ is the variance of complex Gaussian noise.
RMSEs for delay and Doppler
are 
$\frac1{K} \sum_{k=1}^K ( \hat \ell_{\rm d} + \hat \epsilon_{\rm t} - t_{\rm d}/T_s )^2$
and
$\frac1{K} \sum_{k=1}^K ( \hat k_{\rm D} + \hat \epsilon_{\rm f} - f_{\rm D}/ \Delta f )^2$.


\begin{table}[t]
    \centering
    \caption{Execution times (ms: milliseconds)}
    \begin{tabular}{ccc}\hline
      coarse estimation  &  2D sinc interpolation & Quadratic interpolation\\\hline
      7.072 (ms)  &  14.135 (ms) & 0.00388 (ms) \\\hline
    \end{tabular}
    \label{tab:exec_time}
\end{table}

As a baseline, the uniform distribution in $[-0.5,0.5]$ gives the RMSE $1/\sqrt{12}=0.288$.
The proposed method 1 achieves an RMSE of approximately 0.0061 in delay estimation 
and 0.0676 in Doppler estimation at SNR 30 dB.
Algorithm 2 gives an RMSE of 0.0198 in delay estimation, 
while RMSE of 0.1342 in Doppler estimation. 
The RMSEs of the delay are better than those of the Doppler for both algorithms.
This is because the discrete-time auto-ambiguity function $A_{ss}[\ell, k]$ is sharp in delay and wide in Doppler,
as shown in Fig.~\ref{fig:A_ss}

Figs.~\ref{fig:RMSE_delay} and~\ref{fig:RMSE_Doppler} show that the 2-D sinc interpolation gives very accurate estimates. Compared with the baseline performance, quadratic interpolation reduces
the RMSEs to 6.8\% and 46.5\% in delay and Doppler estimation. 
2D sinc interpolation further reduces the RMSEs 30\% and 50\% in delay and Doppler estimation, compared to the quadratic interpolation.  The accuracy of the delay estimation is extremely high.
This is a major advantage of the proposed method for radar applications.

Table~\ref{tab:exec_time} shows execution times, where the coarse estimation implies 
Step 2 of Algorithms 1 and 2, i.e., computation of the discrete-time ambiguity function and find its maximum value, 
and 2D sinc interpolation and Quadratic interpolation mean Step 3 of Algorithms 1 and 2, respectively.  
The execution time of 2D sinc interpolation is about double the computation of the ambiguity function.
Quadratic interpolation, on the other hand, is done instantaneously.
The computational resources required to solve the minimization problem (\ref{argmin_epsilon_t_f}) are not very large
and its computation time is comparable with the FFT computation. Therefore, Algorithm 1 is promising. 
If the radar system cannot tolerate an increase in computational resources, quadratic interpolation with very low computational costs would be an attractive option. 

We used the default settings for BFGS algorithms for Algorithm 1. The execution time can be 
reduced by arranging the algorithm, for example by the setting of a low number of iterations.
We may use (\ref{epsilon_t}) and (\ref{epsilon_f}) as the initial guess for Algorithm 1, which is effective in reducing the number of iterations. 

\section{Conclusion}
We have given an accurate estimation method of delay and Doppler for a radar system using
time- and frequency-shifted Gaussian pulses. 
The main contribution of this paper is that we use $N_{\rm t} \times N_{\rm f}$ time-frequency domain symbols, which allows us to approximate the absolute value of the ambiguity function of the transmitted signal by the 2D sinc functions. Based on this observation, an accurate delay and Doppler estimation method has been proposed.
As the main purpose of radar is to detect the target with high accuracy of the distance and the speed, the proposed radar system is promising. 
In the proposed method, data symbols and random numbers (or pilot symbols) for radar detection can be transmitted simultaneously. Simultaneous data and target detection is a future challenge.

\newpage


\section*{Acknowledgment}
A part of this work was supported by JSPS KAKENHI Grant Number JP19K12156, JP23H00474, and JP23H01409. 



%\bibliographystyle{IEEEtran}

%\bibliographystyle{IEEEtran}
%\bibliography{mybibliography.bib, gyoseki.bib}


\end{document}